\documentclass[aps,prd,twocolumn,superscriptaddress,floatfix,longbibliography]{revtex4-2}
\usepackage{placeins}

\usepackage{graphicx}      % 插入图片
\usepackage{dcolumn}       % 表格中对齐小数
\usepackage{amsmath}       % 数学增强
\usepackage{amssymb}       % 数学符号
\usepackage{xcolor}        % 颜色支持 (required by hyperref)
\usepackage{hyperref}      % 超链接
\usepackage{bm}            % 加粗数学符号
\usepackage{booktabs}      % 三线表
\usepackage{multirow}      % 表格多行合并
\usepackage{lmodern}\usepackage{microtype}%优化字间距和字符突出，减少断词
\usepackage{subcaption}
\usepackage{caption}
\usepackage{comment}
	\makeatletter
	
	\renewcommand\footnote[1]{%
		\stepcounter{footnote}%
		\protected@xdef\@thefnmark{\thefootnote}%
		\@footnotemark\@footnotetext{#1}%
	}
	\makeatother

\begin{document}

\title{Quasinormal modes of Reissner-Nordström black hole from bound state spectrum}

\author{Bo-Yun Zhou}
\affiliation{Key Laboratory of Atomic and Subatomic Structure and Quantum Control (Ministry of Education), Guangdong Basic Research Center of Excellence for Structure and Fundamental Interactions of Matter, School of Physics, South China Normal University, Guangzhou 510006, China and Guangdong Provincial Key Laboratory of Quantum Engineering and Quantum Materials, Guangdong-Hong Kong Joint Laboratory of Quantum Matter, South China Normal University, Guangzhou 510006, China}
\author{Hao-Yun Ma}
\affiliation{Key Laboratory of Atomic and Subatomic Structure and Quantum Control (Ministry of Education), Guangdong Basic Research Center of Excellence for Structure and Fundamental Interactions of Matter, School of Physics, South China Normal University, Guangzhou 510006, China and Guangdong Provincial Key Laboratory of Quantum Engineering and Quantum Materials, Guangdong-Hong Kong Joint Laboratory of Quantum Matter, South China Normal University, Guangzhou 510006, China}
\author{Jia-Hui Huang}
\email{huangjh@m.scnu.edu.cn} 
% \author{Jia-Hui Huang\thanks{huangjh@m.scnu.edu.cn}}
\affiliation{Key Laboratory of Atomic and Subatomic Structure and Quantum Control (Ministry of Education), Guangdong Basic Research Center of Excellence for Structure and Fundamental Interactions of Matter, School of Physics, South China Normal University, Guangzhou 510006, China and Guangdong Provincial Key Laboratory of Quantum Engineering and Quantum Materials, Guangdong-Hong Kong Joint Laboratory of Quantum Matter, South China Normal University, Guangzhou 510006, China}

\begin{abstract}
We apply a newly proposed bound state method to revisit the computation of gravitational and electromagnetic quasinormal modes (QNMs) for Reissner-Nordström black holes (RNBHs). It is found that the method yields QNM frequencies of high accuracy for low-lying modes with overtones $n\leq3$. The accuracy degrades for higher overtones, however, a homotopy deformation to the potential enables us to compute more high overtone modes reliably. 
It is known that the continued fraction method for computing QNMs of extremal RNBHs differs significantly from that used for nonextremal ones. In contrast, the bound state method demonstrates advantage of simplicity: one only need to account for the definition of tortoise coordinate in the extremal case, and the rest of the computational procedure remains exactly the same as that for the nonextremal case.

\end{abstract}
\maketitle

%========================

\section{Introduction}\label{intro}
Quasinormal modes (QNMs) play important roles in the study of how black holes respond to perturbations, with complex frequencies whose real parts correspond to the oscillation frequency and the imaginary parts to the damping rate. The QNM spectra depend on black hole parameters and serve as a powerful probe of the strong-field regime of general relativity and a key tool for interpreting the ringdown signals of binary black hole mergers in gravitational-wave astronomy~\cite{Berti:2009kk,Berti:2025hly,Dreyer:2003bv}. 
Since the first direct detection of gravitational waves from a binary black hole merger by LIGO in 2015~\cite{LIGOScientific:2016aoc}, the precise measurement of black hole mass, spin, and charge from the ringdown spectrum has become an important means to test the general relativity and verify modified theories. The corresponding master equations governing perturbations of various black hole spacetimes have been derived within general relativity and modified theories of gravity~\cite{Regge:1957td,Zerilli:1970se,Teukolsky:1973ha,Cano:2020cao,Cano:2021myl,Wagle:2021tam,Pierini:2021jxd,Pierini2022}.

Over the past four decades, researchers have developed several methods to compute black hole QNMs. One of the well-established methods is Leaver's continued fraction method which requires a careful analysis of the singular points of the master equation, as well as asymptotic solutions at the event horizon and spatial infinity~\cite{Leaver:1985ax,Leaver:1986gd}. 
In 1990, Leaver extended the continued fraction method to the calculation of QNMs for nonextremal RNBH~\cite{Leaver:1990zz}. In this approach, the wave function is expanded as a power series at the horizon, transforming the wave equation into a four-term recurrence relation for the expansion coefficients; by introducing a Gaussian elimination step, a standard three-term recurrence relation is obtained and the QNM frequencies are obtained by a continued fraction algorithm.
However, the method breaks down in the extremal RN limit $(Q=M)$, where the inner and outer horizons merge and the wave equation develops a confluent irregular singularity at the horizon. In this case, the radius of convergence of the conventional power series expansion around the event horizon shrinks to zero, so no valid recurrence relation can be constructed. 
To tackle the QNM calculation for extremal RNBH, Onozawa et al. substantially improved the continued fraction method and uncovered a striking coincidence: in the extremal limit, the QNM frequencies of gravitational perturbations with multipole index $\ell$ coincide exactly with those of electromagnetic perturbations with index $\ell-1$~\cite{Onozawa:1995vu}.

Besides the continued fraction method, researchers have also developed other approaches for computing QNMs of the RNBHs, such as the semi-analytic Wentzel–Kramers–Brillouin (WKB) approximation~\cite{Kokkotas:1988fm} and the phase-amplitude formula proposed by Andersson~\cite{Andersson1993,Andersson:1992scr,Froeman:1992gp}. The WKB method can yield approximate analytical results, but loses accuracy and even breaks down entirely when computing high overtones. In order to mitigate this shortcoming of the WKB method, many researchers have been continuously refining it~\cite{Kokkotas:1988fm,Dunham:1932zz,Schutz:1985km,Iyer:1986np,Iyer:1986nq,Seidel:1989bp,Guinn:1989bn,Kokkotas:1993ef,Matyjasek:2017psv,Konoplya:2019hlu}. 

Another interesting analytic method, the bound state method, was introduced by Mashhoon et al. in 1980s~\cite{Ferrari:1984ozr,Ferrari:1984zz,Blome:1981azp}.  It rests on a profound physical correspondence: the QNM frequencies of a potential barrier can be related to the bound-state energies of the corresponding inverted potential well through an analytic continuation of the relevant parameters. This transforms a scattering problem with outgoing boundary conditions into a standard quantum mechanical bound-state problem, improving the stability of the calculation and providing a fresh perspective on the physical nature of QNMs.
However, this method has historically been limited to potentials whose bound-state spectra are known analytically in closed form, such as the P\"oschl-Teller(PT)~\cite{Poschl:1933zz,Churilova:2021nnc} or Eckart~\cite{Eckart:1930zza} potentials, which are used to locally approximate the effective potential of a black hole near its extreme. Extensions of the application can also be found in~\cite{Zaslavsky:1991ug,Galtsov:1991nwq,Sulejmanpasic:2016fwr,Hatsuda:2019eoj,Matyjasek:2019eeu,Berti:2022xfj,Cheung:2021bol,Cardoso:2024mrw,Chen:2026ehv}.
A significant step toward lifting this restriction was taken by V\"olkel, who proposed computing the bound state energies numerically and constructing a Taylor expansion in model parameters for analytic continuation. This approach was validated for the analytically solvable PT potential and Breit-Wigner (BW) potential, as well as the mixed (PT+BW) potential which has no analytical bound state spectra \cite{Volkel:2022ewm,Volkel:2025lhe}.

In a recent work \cite{Ma:2026bxb}, we have proposed a new method that enables us to calculate QNM frequencies from bound state energies and validated its effectiveness by the calculation of various QNM frequencies for Schwarzschild black holes.
In this method, by a two-step coordinate transformation where an auxiliary real parameter $\alpha$ is introduced, a QNM problem is mapped to a bound state problem, whose eigenvalues $E_n(\alpha)$ are inversely mapped to the QNM frequencies $\omega_n$ via analytic continuation in $\alpha$ from \(E_{n}(\alpha)\).

In this paper, we apply the new bound state method to compute the QNM frequencies of gravitational and electromagnetic perturbations of nonextremal and extremal RNBHs, and in particular, demonstrate the unique advantages of the new method in calculating QNM frequencies for nonextremal and extremal RNBH.

This paper is organized as follows. Section~\ref{sec:method} introduces the single-parameter bound state method for RNBH. Section~\ref{sec:result} presents the results of the QNM frequencies for gravitational and electromagnetic perturbations of the nonextremal and extremal RNBHs, and demonstrates the effect of homotopy deformation on the calculation of higher overtones. Section~\ref{sec:conclusion} is devoted to the conclusion. In the numerical calculation throughout this work we use units in which $G = c = M= 1$.

\section{Linear perturbation equations and method}\label{sec:method}

The development of perturbation theory for RNBH dates back to the 1970s. Zerilli and Moncrief first derived the linear perturbation equations of the RNBH, showing that they decouple into two types of independent wave equations, axial (odd-parity) and polar (even-parity) equations, and that the QNM spectra of the two types are completely identical (isospectral)~\cite{Zerilli:1970se}. Here we just focus on the odd-parity Zerilli-Moncrief equations with multipole index $\ell$ for $Z_{1}^{(-)}, Z_{2}^{(-)}$ ~\cite{Moncrief:1974gw,Chandrasekhar1998,Gunter1980ASO}:

\begin{equation}\label{eq:master}
 \frac{d^{2}}{dx^{2}} Z_{i}^{(-)} + \left[\omega_{}{^2}-V_i(x)\right] Z_{i}^{(-)} =0,
\end{equation}
where
\begin{equation}\label{eq:potential}
	V_i=\frac{\Delta}{r^5}\left[l(l+1)r-q_i+\frac{4Q^2}{r}\right],
\end{equation}
\begin{equation}
	\Delta=r^2-2Mr+Q^2=(r-r_+)(r-r_-),
\end{equation}
\begin{equation}
	r_{\pm}=M\pm\sqrt{M^2-Q^2},
\end{equation}
\begin{equation}
	q_1=3M-\sqrt{9M^2+4Q^2(l-1)(l+2)},
\end{equation}
\begin{equation}
	q_2=3M+\sqrt{9M^2+4Q^2(l-1)(l+2)},
\end{equation}
\begin{equation}
	x=r+\frac{r_{+}^2}{r_+-r_-}\ln(r-r_+)-\frac{r_{-}^2}{r_+-r_-}\ln(r-r_-),
\end{equation}
and $i=1,2$.
Note that $x$ is the tortoise coordinate 
and $\omega_{}$ is the QNM frequencies we want to calculate from the equation. $r_+$ and $r_-$ are the outer and inner horizons of the RNBH. $Z_1^{-}$ and $Z_2^{-}$ describe the odd-parity electromagnetic and gravitational perturbations, respectively. For extremal RNBHs ($Q=M$), the wave equations have a different singularity structure and will be discussed later~\cite{Onozawa:1995vu}.

Consider a coordinate transformation $x\rightarrow -ix$, which maps the QNM problem to a bound state problem~\cite{Ferrari:1984ozr}. Under this transformation, Eq.\eqref{eq:master} becomes
\begin{equation}\label{eq:BS equation}
	\frac{d^2}{dx^2}Z_{i} + \left[E - V_i^{inv}(-ix)\right]Z_{i} = 0,
\end{equation}
where $E=-\omega^2$, the inverted potential is defined as $V_i^{inv}(-ix)=-V_{i}(-ix)$. Next, we introduce a real scaling parameter $\alpha$ into the potential in \ref{eq:BS equation} via a substitution $-ix\to\alpha x$~\cite{Ma:2026bxb}. For any real value $\alpha \neq 0$, Eq.\eqref{eq:BS equation} becomes 

\begin{equation}
	\frac{d^2}{dx^2}Z_{i} + \left[E - V_i^{inv}(\alpha x)\right]Z_{i}= 0.
	\label{eq:bound}
\end{equation}
where $V_i^{inv}(\alpha x)=-V_{i}(\alpha x)$. This equation defines a standard bound state eigenvalue problem whose energies $E_n(\alpha)$ can be computed numerically. 
After getting the bound state energies, we analytically continue the results to the target point $\alpha =-i$, and the QNM frequencies can be obtained as $\omega_n = \sqrt{-E_n(\alpha =-i)}$. 

In practice, we numerically compute $\{E_n(\alpha_j)\}$ on a uniform grid of real values $\alpha$ centered at $\alpha_0$ with step-size
$h$ by using the shooting method, fit the data to a numerical representation of $E_n(\alpha)$ using diagonal or subdiagonal Padé approximants, and then perform the analytic continuation of $E_n(\alpha)$ to obtain the QNM frequencies.

Although numerical results of electromagnetic and gravitational QNM frequencies for nonextremal RNBHs can be found in \cite{Leaver:1990zz,Andersson1993},
in order to show the relative errors between the results computed with bound state method and Leaver's continued fraction method, we calculate the QNM frequencies with Leaver's method again and retain 15 decimal places.

\section{results}\label{sec:result}
In this section, we show the numerical results of the electromagnetic and gravitational QNM frequencies obtained with the newly proposed bound state method.
Unless otherwise stated, in the numerical calculation for both nonextremal and extremal RNBHs, 
we use a 31-point grid centered at $\alpha_0=0.5$ to compute the bound state energies $E_{n}(\alpha)$, with $\alpha$ ranging from 0.2 to 0.8 in steps of 0.02. The bound states are computed to 30 decimal places. Diagonal or subdiagonal Padé approximants are also used to approximate $E_{n}(\alpha)$ and to analyze the singularities of $E_{n}(\alpha)$.

\subsection{Nonextremal RNBHs}
In Table~\ref{tab:QNM_grav} and Table~\ref{tab:QNM_elec}, we present various values of the gravitational and electromagnetic QNM frequencies for the nonextremal RNBH, respectively. Comparisons between the results with bound state method and that with Leaver's continued fraction method are also shown in these tables.
The relative errors ($\delta^{r(i)}=|\omega^{r(i)}_{\rm \text{BS}}-\omega^{r(i)}_{\rm Leaver}|/|\omega^{r(i)}_{\rm Leaver}|$) of the real and imaginary parts are shown in Table~\ref{tab:grav_wucha} and Table~\ref{tab:elec_wucha}, respectively.
\begin{table*}[h]
	\caption{\label{tab:QNM_grav} The gravitational QNM frequencies with $l = 2$ for RNBH with various values of $Q$, which are computed with bound state (BS) method and Leaver's method, respectively.}
	\begin{ruledtabular}
		\begin{tabular}{cccccc}
			  $Q$ & Method & $n=0$ & $n=1$ &$n=2$& $n=3$ \\ \hline
			  0.2 &   BS  &0.37474+0.08907 i & 0.34783+0.27423 i & 0.30223+0.47875 i & 0.25081+0.70830 i\\
			      &Leaver &0.37474+0.08907 i & 0.34783+0.27423 i &0.30222+0.47874 i &0.25268+0.70568 i\\ \hline
		      0.4 &   BS  &0.37844+0.08940 i & 0.35172+0.27512 i & 0.30643+0.47994 i &0.25484+0.71113 i\\
		          &Leaver &0.37844+0.08940 i & 0.35173+0.27512 i &0.30642+0.47994 i &0.25695+0.70692 i\\ \hline
			  0.6 &   BS  &0.38622+0.08981 i & 0.36017+0.27615 i & 0.31589+0.48088 i &0.26442+0.71032 i\\
			      &Leaver &0.38622+0.08981 i & 0.36017+0.27615 i &0.31588+0.48088 i &0.26704+0.70704 i\\ \hline
			  0.8 &   BS  &0.40122+0.08964 i & 0.37690+0.27494 i & 0.33517+0.47656 i &0.28619+0.69818 i \\
			      &Leaver &0.40121+0.08964 i & 0.37690+0.27494 i &0.33517+0.47656 i &0.28773+0.69737 i\\ \hline
			  0.99 &  BS  &0.42930+0.08427 i & 0.40352+0.25701 i & 0.35392+0.44358 i &0.28899+0.65387 i \\
			      &Leaver &0.42930+0.08427 i & 0.40352+0.25701 i &0.35391+0.44358 i &0.28864+0.65391 i \\
		\end{tabular}
	\end{ruledtabular}
\end{table*}

\begin{table*}[h]
	\caption{\label{tab:QNM_elec} The electromagnetic QNM frequencies with $l = 2$ for RNBH with various values of $Q$, which are computed with bound state method and Leaver's method, respectively.}
	\begin{ruledtabular}
		\begin{tabular}{cccccccc}
			$Q$ & Method & $n=0$ & $n=1$ &$n=2$&$n=3$&$n=4$  \\ \hline
			0.2 &   BS  &0.46296+0.09537 i & 0.44218+0.29173 i & 0.40726+0.50304 i &0.36908+0.73181 i &0.33286+0.97161 i\\
			&Leaver &0.46296+0.09537 i & 0.44218+0.29173 i &0.40726+0.50304 i &0.36907+0.73184 i&0.33549+0.97332 i\\ \hline
			0.4 &   BS  &0.47992+0.09644 i & 0.46000+0.29469 i & 0.42648+0.50715 i &0.38960+0.73630 i &0.35645+0.97756 i\\
			&Leaver &0.47992+0.09644 i & 0.46000+0.29469 i &0.42648+0.50715 i &0.38960+0.73631 i&0.35688+0.97780 i\\ \hline
			0.6 &   BS  &0.51201+0.09802 i & 0.49375+0.29891 i & 0.46293+0.51261 i&0.42858+0.74149 i&0.39395+0.98534 i \\
			&Leaver &0.51201+0.09802 i & 0.49376+0.29891 i &0.46293+0.51261 i &0.42859+0.74149 i&0.39762+0.98195 i\\ \hline
			0.8 &   BS  &0.57013+0.09906 i & 0.55499+0.30105 i & 0.52912+0.51315 i&0.49939+0.73746 i&0.47091+0.97219 i \\
			&Leaver &0.57013+0.09906 i & 0.55499+0.30105 i &0.52912+0.51315 i &0.49939+0.73746 i&0.47139+0.97170 i\\ \hline
			0.99 &  BS  &0.69275+0.08864 i &0.67865+0.26750 i &0.65095+0.45118 i&0.61080+0.64321 i&0.56084+0.84742 i \\
			&Leaver &0.69275+0.08864 i &0.67865+0.26750 i &0.65095+0.45118 i &0.61080+0.64321 i&0.56084+0.84742 i\\ 
		\end{tabular}
	\end{ruledtabular}
\end{table*}

\begin{table*}[h]
	\caption{\label{tab:grav_wucha} Relative errors of the real and imaginary parts of gravitational QNMs given in Table \ref{tab:QNM_grav}. }
	\begin{ruledtabular}
		\begin{tabular}{ccccccccc}
			$Q$ & $ \delta^{r}$ & $\delta^{i}$ & $\delta^{r}$ & $\delta^{i}$ & $ \delta^{r}$ & $\delta^{i}$& $ \delta^{r}$ & $\delta^{i}$\\ 
			&$n=0$	&$n=0$	&$n=1$	&$n=1$	&$n=2$	&$n=2$&$n=3$	&$n=3$	\\ \hline
			0.2 & $3\times 10^{-10}$  &$2\times 10^{-09}$  & $2\times 10^{-07}$ & $1\times 10^{-07}$ & $1\times 10^{-05}$ &$2\times 10^{-05}$& $7\times 10^{-03}$ &$4\times 10^{-03}$ \\
			0.4 & $2\times 10^{-09}$ &$5\times 10^{-09}$  & $2\times 10^{-07}$ & $3\times 10^{-08}$ & $9\times 10^{-06}$ &$1\times 10^{-07}$ & $8\times 10^{-03}$ &$6\times 10^{-03}$\\
			0.6 & $5\times 10^{-10}$ &$2\times 10^{-09}$  & $1\times 10^{-07}$ & $2\times 10^{-07}$ & $1\times 10^{-05}$ &$1\times 10^{-06}$ & $1\times 10^{-02}$ &$5\times 10^{-03}$\\
			0.8 & $8\times 10^{-11}$ &$1\times 10^{-10}$  & $6\times 10^{-08}$ & $6\times 10^{-08}$ & $2\times 10^{-06}$ &$1\times 10^{-06}$ & $5\times 10^{-03}$ &$1\times 10^{-03}$\\
			0.99 & $2\times 10^{-10}$ &$1\times 10^{-09}$  & $3\times 10^{-08}$ & $1\times 10^{-07}$ & $2\times 10^{-05}$ &$9\times 10^{-07}$& $1\times 10^{-03}$ &$6\times 10^{-05}$ \\
		\end{tabular}
	\end{ruledtabular}
\end{table*}

\begin{table*}[h]
	\caption{\label{tab:elec_wucha} Relative errors of the real and imaginary parts of electromagnetic QNMs given in Table \ref{tab:QNM_elec}.}
	\begin{ruledtabular}
	\begin{tabular}{ccccccccccc}
			$Q$ & $ \delta^r$ & $\delta^i$ & $\delta^r$ & $\delta^i$ & $ \delta^r$ & $\delta^i$& $\delta^r$ & $\delta^i$& $\delta^r$ & $\delta^i$\\ 
			&$n=0$	&$n=0$	&$n=1$	&$n=1$	&$n=2$	&$n=2$	&$n=3$	&$n=3$&$n=4$	&$n=4$\\ \hline
			0.2 & $8\times 10^{-14}$ &$4\times 10^{-12}$  & $6\times 10^{-10}$ & $2\times 10^{-10}$ & $7\times 10^{-08}$ &$2\times 10^{-08}$& $8\times 10^{-06}$ &$3\times 10^{-05}$ &$8\times 10^{-03}$&$2\times 10^{-03}$\\
			0.4 & $6\times 10^{-15}$ &$2\times 10^{-12}$  & $2\times 10^{-10}$ & $2\times 10^{-10}$ & $2\times 10^{-08}$ &$1\times 10^{-08}$& $2\times 10^{-05}$ &$2\times 10^{-05}$ &$1\times 10^{-03}$&$2\times 10^{-04}$\\
			0.6 & $9\times 10^{-14}$ &$6\times 10^{-14}$  & $2\times 10^{-10}$ & $4\times 10^{-10}$ & $3\times 10^{-08}$ &$8\times 10^{-09}$& $2\times 10^{-05}$ &$4\times 10^{-06}$&$9\times 10^{-03}$&$3\times 10^{-03}$ \\
			0.8 & $6\times 10^{-14}$ &$1\times 10^{-13}$  & $1\times 10^{-11}$ & $5\times 10^{-11}$ & $5\times 10^{-10}$ &$4\times 10^{-10}$& $1\times 10^{-07}$ &$3\times 10^{-06}$&$1\times 10^{-03}$&$5\times 10^{-04}$ \\
			0.99 & $9\times 10^{-14}$ &$7\times 10^{-13}$  & $2\times 10^{-11}$ & $3\times 10^{-11}$ & $3\times 10^{-10}$ &$2\times 10^{-09}$& $2\times 10^{-08}$ &$2\times 10^{-08}$&$9\times 10^{-06}$ &$3\times 10^{-06}$ \\
		\end{tabular}
	\end{ruledtabular}
\end{table*}

From Table ~\ref{tab:QNM_grav} and Table~\ref{tab:grav_wucha}, we can see that for gravitational QNMs with $\ell=2$, the bound state method can yield results of high or good accuracy 
and achieve a relative error of $\lesssim 10^{-5}$ when $n\leq 2$; when $n=3$, the bound state method can also yield reliable results. We also find that this method fails for higher overtones, and increasing the numerical precision from 30 to 35 decimal digits and the number of $\alpha$-sampling points from 31 to 51 improves the accuracy of the QNM frequencies for all modes from $n=0$ to $n=3$, but still fails to produce reliable results for $n\geq4$.

For electromagnetic QNM frequencies (Table~\ref{tab:QNM_elec} and Table~\ref{tab:elec_wucha}), the results exhibit a similar behavior, and the method fails to produce reliable results for $n\geq5$. The reason for the failure is that the singularity of the bound state energy $E_n(\alpha)$ lies inside the circle of analytic continuation, which is similar to the Schwarzschild case \cite{Ma:2026bxb}. 
As $Q$ increases, the accuracy of the QNM frequencies improves, most notably for modes with $n\geq 2$. A physical interpretation of this is that the potential well becomes deeper as
$Q$ increases, and the distribution of the wavefunctions of higher overtone modes are more strongly affected by the depth of the potential well.   
Compared with the gravitational case, the electromagnetic QNM frequencies can be computed reliably up to $n=4$ with the bound state method. This can also be attributed to the fact that the potential well for electromagnetic perturbation is deeper than that for the gravitational perturbation \cite{Volkel:2025lhe,Li:2026ptw}.

\subsection{Effects of homotopy deformations}\label{sec:homotopy}

A homotopy deformation procedure was proposed and demonstrated to be useful for calculation of QNM frequencies of higher overtones in the Schwarzschild case~\cite{Ma:2026bxb}.
In this subsection, we will show that it is also useful for computing QNM frequencies of higher overtones for the RNBH. 
 
The homotopy deformation function can be chosen as follows:
	\begin{equation}
		\mathcal{H}(\alpha,\lambda) = \left(\lambda-1\right)\alpha^2+\lambda.
		\label{eq:deepening}
	\end{equation}
	Then, the deformed potential well is
	\begin{equation}
	V_{\rm \lambda}(\alpha x)=\mathcal{H}(\alpha,\lambda) V_{\rm inv}(\alpha x).
	\end{equation}
When $\lambda=1$, it reduces to the original inverted potential $V_{\rm inv}(\alpha x)$. As $\lambda$ increases, the potential is continuously scaled by $\mathcal{H}(\alpha,\lambda)$ for real $\alpha$. The parameter $\lambda$ thus serves as a homotopy parameter, continuously deforming the potential well into a deeper one. Since $\mathcal{H}(\alpha=-i,\lambda) = 1$ holds for all $\lambda$, the black hole QNM problem is exactly restored at the target point $\alpha=-i$ for all $\lambda$.

\begin{table}[htbp]
	\caption{\label{tab:QNM_grav_deep} QNM frequencies of gravitational perturbations ($l = 2$, $n=4$) for RNBH, with homotopy parameter $\lambda =4$.} 
	\begin{ruledtabular}
		\begin{tabular}{cccccc}
			$Q$   & $\omega_{\rm Origin}$  & $\omega_{\rm Homotopy}$ & $\omega_{\rm CFM}$   \\ \hline
				0.2 &0.21407+0.93017i &0.20936+0.94965i & 0.20852+0.94744i  \\ \hline
				0.4 &0.21490+0.93034i &0.21279+0.94989i & 0.21249+0.94868i   \\\hline
				0.6 &0.18004+0.92458i &0.22290+0.94914i & 0.22235+0.94776i  \\ \hline
				0.8 &0.24502+0.91551i &0.24211+0.93297i & 0.24219+0.93184i  \\ \hline
				0.99 &0.25366+0.89787i &0.21896+0.88914i & 0.22172+0.88920i  \\ 
			\end{tabular}
		\end{ruledtabular}
	\end{table}
	
	\begin{table}[htbp]
		\caption{\label{tab:QNM_elec_deep} QNM frequencies of electromagnetic perturbations ($l = 2$, $n=5$) for RNBH, with homotopy parameter $\lambda =4$.} 
		\begin{ruledtabular}
			\begin{tabular}{cccccc}
				$Q$  & $\omega_{\rm Origin}$  & $\omega_{\rm Homotopy}$ & $\omega_{\rm CFM}$   \\ \hline
				0.2 &0.38788+1.06560i &0.30911+1.22185i & 0.30842+1.22150i  \\ \hline
				0.4 &0.18926+0.53376i &0.33108+1.22633i & 0.33034+1.22596i   \\\hline
				0.6 &0.49227+1.16054i &0.37283+1.22939i & 0.37219+1.22897i  \\ \hline
				0.8 &0.44175+1.20965i &0.44839+1.21234i & 0.44740+1.21209i  \\ \hline
				0.99 &0.50576+1.06676i &0.50683+1.06637i & 0.50623+1.06661i  \\ 
			\end{tabular}
		\end{ruledtabular}
		\end{table}
In Table~\ref{tab:QNM_grav_deep} and Table~\ref{tab:QNM_elec_deep}, we show several gravitational and electromagnetic QNM frequencies of higher overtones, respectively.
$\omega_{\rm Origin}$ and $\omega_{\rm Homotopy}$ denote the QNM frequencies obtained without and with the homotopy deformation, respectively. $\omega_{\rm CFM}$ are the reference values calculated by the continued fraction method \cite{Leaver:1990zz}. 
We can see that after applying a homotopy deformation ($\lambda =4$) to the potential wells, reliable QNM frequencies can be obtained up to $n=4$ for gravitational perturbations, and up to $n=5$ for electromagnetic perturbations. In fact, we also find that when $Q>0.8$ reliable electromagnetic QNM frequencies can be computed up to $n = 6$. 

\begin{figure}[htbp] 
	\centering   %图片居中
	\includegraphics[width=1\columnwidth]{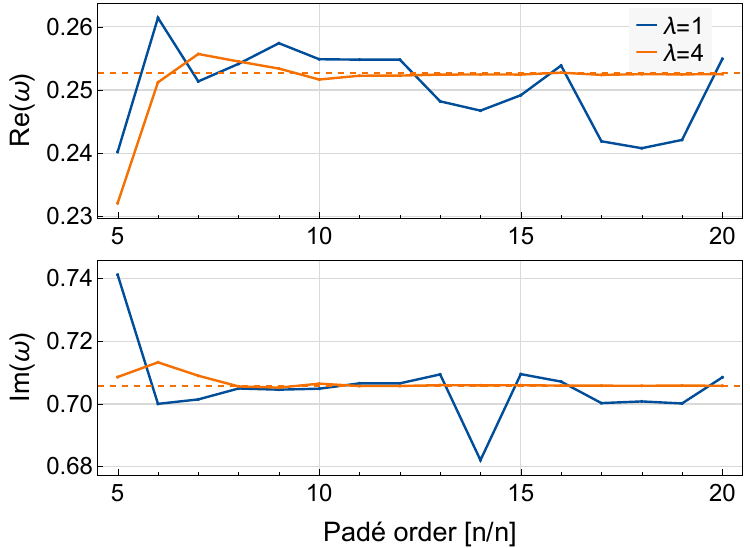}
	\caption{Convergence curves of a gravitational QNM frequency
		($l=2$, $Q=0.2$, $n = 3$) with increasing Pad\'e order. Dashed lines represent reference values from
		Leaver’s method.}
	\label{fig}  
\end{figure}
In Fig.~\ref{fig}, we demonstrate the effect of the homotopy deformation on the convergence of a representative QNM frequency. 
Here we use a 41-point grid centered at $\alpha_0=0.5$ to compute $E_{n}(\alpha)$, with $\alpha$ ranging from 0.1 to 0.9 in steps of 0.02. 
One can see that, prior to a homotopy deformation to the potential well ($\lambda = 1$), the QNM frequency does not converge as the Pad\'e order increases. In contrast, after applying a suitable deformation ($\lambda = 4$), both the real and imaginary parts converge reliably from the 11th-order Pad\'e approximant onward.

However, $\lambda$ is not necessarily the larger the better, it is found that for $\lambda > 8$, the accuracy of the QNM frequencies is lower than that for $\lambda= 4$.
By studying the singularities of the bound state energies $E_{n}(\alpha)$ of the RNBH, it is found that similar to the SchBH case~\cite{Ma:2026bxb}, these singularities
also gradually approach the target point $\alpha =-i$ as $\lambda$ increases, which explains why an excessively large  $\lambda$ leads to a decline in the accuracy of the QNM frequencies.

\subsection{Extremal RNBH}
When applying Leaver's continued fraction method to the calculation of QNM frequencies of extremal RNBH, a special treatment is needed~\cite{Onozawa:1995vu}. This is because 
the inner and outer horizons merge for the extremal RNBH, the wave equation of the perturbations develops a confluent irregular singularity at the horizon, the radius of convergence of the conventional power series expansion around the horizon shrinks to zero, and no valid recurrence relation can be constructed. In~\cite{Onozawa:1995vu}, the author carefully analyzed the structure of the wave equation and developed a modified continued fraction method where the solution is expanded around a suitable ordinary point of the wave equation.   
For gravitational or electromagnetic QNMs, a complicated way was also developed to obtain the recurrence relations for even-numbered and odd-numbered overtones, respectively.

Here, we demonstrate that the bound state method proposed in \cite{Ma:2026bxb} allows for a straightforward computation of QNM frequencies for extremal RNBHs. The procedure is essentially identical to that for the nonextremal case, and one only need to account for the redefinition of the tortoise coordinate, whose definition is
\begin{equation}\label{eq:new tortoise}
	x= \int \frac{r^2}{\Delta}\,dr =r-\frac{M^2}{r-M}+2M\ln(r-M).
\end{equation}

In Table \ref{tab:QNM_extrem}, we present the QNM frequencies of the extremal RNBH calculated with the bound state method proposed in \cite{Ma:2026bxb}, and we also compare 
our results with that obtained with continued fraction method in~\cite{Onozawa:1995vu}.  
\begin{table*}[htbp]
	\caption{\label{tab:QNM_extrem}QNM frequencies of the gravitational and electromagnetic perturbations for extremal RNBH. The reference results are from~\cite{Onozawa:1995vu}.}.
	\begin{ruledtabular}
		\begin{tabular}{cccccc}
			Type	&$\ell$ & Method & $n=0$ & $n=1$ & $n=2$  \\ \hline
			&2 &   BS  &0.4313408004+0.0834603150 i & 0.4045235541+0.2549843642 i & 0.3534012455+0.4413746578 i \\
			& &Onozawa &0.43134+0.083460 i & 0.40452+0.25498 i &0.35340+0.44137 i \\ 
			Grav &3 &   BS  &0.7043040115+0.0859733739 i & 0.6880422456+0.2599232479 i & 0.6562419003+0.4400688689 i \\
			& &Onozawa &0.70430+0.085973 i & 0.68804+0.25992 i &0.65624+0.44007 i \\ 
			&4 &   BS  &0.9657626042+0.0870013210 i & 0.9538122060+0.2621199348 i & 0.9302024350+0.4406437804 i \\
			&  &Onozawa &0.96576+0.087001 i & 0.95381+0.26212 i &0.93020+0.44064 i \\ \hline
			&1 &  BS  &0.4313408004+0.0834603150 i & 0.4045235541+0.2549843642 i & 0.3534012455+0.4413746578 i \\
			Elec   &2 &  BS  &0.7043040115+0.0859733739 i & 0.6880422456+0.2599232479 i & 0.6562419003+0.4400688689 i \\ 
			&3 &  BS  &0.9657626042+0.0870013210 i & 0.9538122060+0.2621199348 i & 0.9302024350+0.4406437804 i \\
		\end{tabular}
	\end{ruledtabular}
\end{table*}

It is observed that the $n\leq 2$ QNM frequencies obtained with the bound state method agree exactly with the continued-fraction values to five decimal places.
Our numerical results also show the isospectrality between electromagnetic QNMs with multipole index $\ell$ and gravitational QNMs with multipole index $\ell+1$.
In fact, this isospectrality manifests itself not only in the QNM frequencies, but also in the bound state energies of the electromagnetic and
gravitational perturbations.

 \section{Conclusion}\label{sec:conclusion}
 
 In this work, using a newly proposed bound state method, we revisit the computation of electromagnetic and gravitational QNMs for nonextremal and extremal RNBHs. 
 The method yields QNM frequencies of high accuracy for low-lying modes with $n\lesssim l$. A homotopy deformation enables the computation of several higher overtones.
We do not provide a detailed discussion of the singularities of the bound state energies\(E_{n}(\alpha)\) 
 since it is similar to that for the SchBHs \cite{Ma:2026bxb}.
 
Particularly, compared with the traditional continued fraction method where extremal RNBHs demand a separate and fundamentally different treatment from that of nonextremal ones,
the newly proposed method demonstrates its simplicity and universality: apart from the definition of tortoise coordinate, all other procedure is identical for both extremal and nonextremal RNBHs.

This work exemplifies that the newly proposed bound state method is effective for high accurate calculation of low-lying QNM frequencies, and is particular useful for spherically symmetric black holes with complicate singularity structures. 

Although the homotopy deformation used in this work can enable us to compute several higher overtone modes, it remains powerless for more higher overtones. 
Thus, an interesting further direction is to explore whether there is a deformation which could systematically resolve the limitation on the computation of higher overtones
with the method. It will also be interesting to apply the method to calculate QNM frequencies of other spherically symmetric black holes.

%\begin{acknowledgments}
%This is a preliminary preprint; we
%welcome any comments to help improve this work.
%\end{acknowledgments}

\bibliography{reference}

\clearpage
%\bibliography{wkb}

\end{document}